\pdfoutput=1
\documentclass[graybox]{svmult}
\usepackage{mathptmx}      
\usepackage{helvet}         
\usepackage{courier}       
\usepackage{makeidx}         
\usepackage{graphicx}        
\usepackage{multicol}       
\usepackage[bottom]{footmisc}
\usepackage{amsmath,amssymb,amsfonts}       
\usepackage[colorlinks=true,linkcolor=black,citecolor=magenta,urlcolor=blue,filecolor=black]{hyperref}
\makeindex

\newcommand{\eq}[2]{\begin{equation} #1 \label{#2} \end{equation}}
\DeclareMathOperator{\extdm}{d}
\newcommand{\extd}{\extdm \!}

\newcommand{\whatever}{I_0} % this is the name for the numerical factor in the Killing norm, whose origin is the freedom to choose the integration constant in Q(X)

\begin{document}
\newcommand{\mytitle}{Minkowski and (A)dS ground states in general 2d dilaton gravity}
\title*{\mytitle}
% Use \titlerunning{Short Title} for an abbreviated version of
% your contribution title if the original one is too long
\author{Daniel Grumiller, Martin Laihartinger and Romain Ruzziconi}
% Use \authorrunning{Short Title} for an abbreviated version of
% your contribution title if the original one is too long
\institute{Daniel Grumiller \at Institute for Theoretical Physics, TU Wien, Wiedner Hauptstrasse 8-10, A-1040 Vienna, Austria, \email{grumil@hep.itp.tuwien.ac.at}
\and Martin Laihartinger \at Institute for Theoretical Physics, TU Wien, Wiedner Hauptstrasse 8-10, A-1040 Vienna, Austria, \email{e1608104@student.tuwien.ac.at} \and Romain Ruzziconi \at Institute for Theoretical Physics, TU Wien, Wiedner Hauptstrasse 8-10, A-1040 Vienna, Austria, \email{romain.ruzziconi@tuwien.ac.at}}
\maketitle

\abstract{
We reorganize the derivative expansion of general (power-counting non-renormalizable) 2d dilaton gravity such that the mass function is integrable. As an example, we consider a three-parameter family of models and provide conditions on the parameters such that the ground state is either Minkowski, Rindler, or (A)dS.
}

\section{Introduction to general 2d dilaton gravity}
\label{sec:1}

Two is the lowest spacetime dimension with enough room for space and time. Consequently, it is the lowest dimension that allows for lightcones, horizons, and black holes. Moreover, two is the lowest dimension permitting non-trivial intrinsic curvature and topology. For all these reasons, two-dimensional (2d) theories of gravity serve as minimal toy models to elucidate conceptual aspects of classical and quantum gravity, including black hole evaporation and holographic correspondences.

Since Einstein gravity does not exist in 2d, one has to resort to some alternative gravity theory. All known approaches\footnote{
This includes intrinsic models \cite{Jackiw:1985je,Teitelboim:1983ux,Callan:1992rs}, higher curvature/torsion models \cite{Katanaev:1986wk}, gravity as linear \cite{Isler:1989hq,Chamseddine:1989yz,Verlinde:1991rf} or non-linear gauge theory \cite{Ikeda:1993fh,Schaller:1994es}, the target-space action of non-critical strings in 2d \cite{Mandal:1991tz,Elitzur:1991cb,Witten:1991yr,Dijkgraaf:1992ba}, a scaling limit of gravity in $2+\epsilon$ dimensions \cite{Mann:1992ar,Lemos:1993py,Grumiller:2007wb} and various dimensional reductions to 2d, see \cite{Grumiller:2002nm} and additional refs.~therein.} lead to the same class of theories (possibly after some field redefinitions and/or integrating in/out auxiliary fields), namely 2d dilaton gravity, a scalar-tensor theory. In its most general version, the bulk action  
\eq{
I[g_{\mu\nu},\,X] =- \frac{\kappa}{4\pi}\,\int\extd^2x\sqrt{-g}\,\Big(XR-2{\cal V}\big(X,\,-(\partial X)^2\big)\Big) 
}{eq:1}
contains an arbitrary function $\cal V$ of two variables, the dilaton $X$ and its kinetic term $(\partial X)^2$. We adapted signs consistently with \cite{Grumiller:2021cwg}. Special cases often considered in the literature are the restriction to power-counting renormalizable models \cite{Odintsov:1991qu,Russo:1992yg} or to potentials that are independent from the kinetic term \cite{Louis-Martinez:1993bge,Witten:2020ert}. In our work, we do not impose such restrictions.

While the action \eqref{eq:1} has appeared, e.g., in \cite{Strobl:1999wv,Grumiller:2002nm,Grumiller:2002md,Kunstatter:2015vxa,Takahashi:2018yzc} (also known as ``kinetic gravity braiding'' \cite{Deffayet:2010qz} or ``Horndeski theory'' \cite{Horndeski:1974wa}), besides the class of dilaton scale invariant models studied in detail in \cite{Grumiller:2021cwg} no charting of the full model space was attempted yet.

In this proceedings contribution, we make the first steps in this direction. In section \ref{sec:2} we review how to obtain all classical solutions, focussing on the linear dilaton sector, up to the integration of a conservation equation. In section \ref{sec:3} we reorganize the way that models are formulated by demanding the conservation equation to be integrable and providing some function therein as input, from which the potential $\cal V$ is then deduced. In section \ref{sec:4} we focus on a three-parameter family as an example and discuss under which conditions the models exhibit ground states that are maximally symmetric (Minkowski, Rindler, or (A)dS). In section \ref{sec:6} we conclude.

\section{Classical solutions}
\label{sec:2}

To discuss the classical solutions of general 2d dilaton gravity \eqref{eq:1}, it is convenient to work in the first order formulation of the theory \cite{Klosch:1995fi,Klosch:1995qv}, whose action 
\eq{
I[\omega,\,e_a,\,X,\,X^a] = \frac{\kappa}{2\pi}\int\Big(X\,\extd\omega + X^a\,\big(\extd e_a + \epsilon_a{}^b\,\omega\wedge e_b\big) + \epsilon \, {\cal V}(X,\,X^c X_c)\Big)
}{eq:g2d11} 
depends on the (dualized) Lorentz connection $\omega$, the zweibein $e_a$, the dilaton $X$, and auxiliary scalar fields $X^a$. We introduced the volume form $\epsilon=\frac12\epsilon^{ab}e_a\wedge e_b$. In light-cone coordinates $(x^+,x^-)$ for the Lorentz indices $a,b$, the Minkowski metric reads as $\eta_{+-}=1$, $\eta_{\pm\pm}=0$. Our sign convention for the Levi-Civit\'a symbol is $\epsilon^\pm{}_\pm=\pm 1$. The equations of motion 
\begin{subequations}\begin{align}
&\extd X + X^- e^+ - X^+ e^- = 0 \label{eq1} \\
&(\extd\pm  \omega ) X^\pm \pm  e^\pm\,\mathcal{V}  = 0 \label{eq2} \\
&\extd \omega + \epsilon\, \frac{\partial \mathcal{V}}{\partial X} = 0  \label{eq3}\\
&(\extd \pm \omega) e^\pm + \epsilon\, \frac{\partial \mathcal{V}}{\partial X^\mp} =0 \label{eq4}
\end{align}\end{subequations}   
are of first order in derivatives. Solving \eqref{eq3} and \eqref{eq1} for $\omega$ and $X^a$ and re-injecting these solutions into the action \eqref{eq:g2d11} recovers the second order action \eqref{eq:1} (see e.g.~\cite{Grumiller:2021cwg} for details). 

The constant dilaton sector of the theory is given by the solutions of \eqref{eq1}-\eqref{eq4} satisfying 
\eq{
{\cal V}(X,\,X^aX_a)=0=X^a
}{eq:cdv} 
These solutions have constant dilaton, constant curvature, and are locally maximally symmetric. They are covered extensively in the literature (see e.g.~\cite{Hartman:2008dq}), so we do not discuss them in this work.

Instead, we focus on the linear dilaton sector of the theory and summarize the solution algorithm discussed in \cite{Grumiller:2021cwg}. Combining \eqref{eq2} and \eqref{eq1} obtains
\begin{equation}
    \extd\,Y - \mathcal{V}(X,\,2Y)\,\extd X = 0 \qquad\qquad Y:=X^+X^-\,.
    \label{stage 1}
\end{equation}
As discussed below, this relation implies Casimir conservation $\extd C=0$ and can be integrated to yield the Casimir function $C(X,\,Y)$. Assuming $X^+ \neq 0$ (essentially without loss of generality) the equation \eqref{eq2} implies 
\begin{equation}
\omega = -\frac{\extd X^+}{X^+} - Z\, \mathcal{V}
\label{stage 4}
\end{equation} 
where $Z = e^+ /X^+$. Similarly, the equation \eqref{eq1} yields
\begin{equation}
e^- = \frac{\extd X}{X^+} + X^- Z
\label{stage 2}
\end{equation} 
and the volume element $\epsilon=- Z\wedge \extd X$. Combining the upper signs \eqref{eq4} and \eqref{eq2} gives
\begin{equation}
\extd Z =\left( Z\wedge \extd X\right)\frac{\partial {\cal V}}{\partial Y} \,.
\end{equation}
Inserting $Z = \extd v\, e^{Q(X)}$ into this equation yields
\begin{equation}\label{EqforQ}
    \frac{\extd Q}{\extd X}=-\frac{\partial {\cal V}}{\partial Y} 
\end{equation} 
which can be formally integrated as
\begin{equation}
    Q(X)= -\int^X\frac{\partial \cal V}{\partial Y} \label{Q solution}
\end{equation} 
by virtue of \eqref{stage 1}. The line element reads as
\begin{equation}
\extd s^2 = 2 e^+ e^- = 2 e^Q\, \extd v \extd X + 2 Y e^{2 Q}\, \extd v^2   \label{solution in metric}
\end{equation} 
where $Y$ is constrained through \eqref{stage 1} and $Q(X)$ is given in \eqref{Q solution}. Introducing the radial coordinate $\extd r = e^Q \extd X$, which can be integrated to give the dilaton as a function of the radius $X(r)$, the line element \eqref{solution in metric} takes Eddington--Finkelstein form,
\begin{equation}
    \extd s^2 = 2 e^{2Q}Y \extd v^2 + 2\extd v\extd r \,.
    \label{solution in metric2}
\end{equation}
The solution \eqref{solution in metric2} is completely determined once the potential $\mathcal{V}$ is provided, up to the integration constant $C$ in \eqref{stage 1} corresponding to the Casimir of the theory.\footnote{The constant coming from the integration of \eqref{EqforQ} is trivial since it can be fixed by a choice of units.} All solutions exhibit at least one Killing vector, namely $\partial_v$.

We highlight three properties of the Casimir $C$: 1.~Physically, it encodes the mass of the solution. 2.~Notably, the function $Q(X)$, in general, depends on the Casimir, except for power-counting renormalizable models. 3.~To obtain explicit solutions, we still need to integrate \eqref{stage 1}. How to achieve this is the subject of the next section.

\section{Integrable conservation equation}
\label{sec:3}

In practice, given a potential $\mathcal{V}$, it might be difficult to integrate the conservation equation \eqref{stage 1} and find an explicit solution for $Y$. Furthermore, it is not a priori obvious to determine which class of potentials $\mathcal{V}$ will allow us to find explicit solutions and obtain a complete classification of the ground states. Instead, we will use another route and start from a class of exact differential equations for \eqref{stage 1}. They are of the form
\begin{equation}
	\label{ede}
	p(X,\,Y) ~ \extd X + q(X,\,Y) ~ \extd Y = 0 
\end{equation} 
where $p(X,\,Y)$ and $q(X,\,Y)$ are functions satisfying the integrablity condition 
\begin{equation}
	\label{side3}
	\frac{\partial p}{\partial Y} = \frac{\partial q}{\partial X} \,.
	\end{equation} 
With this condition, the conservation equation \eqref{ede} can be rewitten in terms of the Casimir $C(X,\,Y)$ as 
\eq{
\extd C = 0 \qquad\textrm{with}\qquad p=\frac{\partial C}{\partial X} \qquad q=\frac{\partial C}{\partial Y}
}{eq:angelinajolie}
which readily can be integrated. Comparing with \eqref{stage 1}, we identify 
	\begin{equation}
	  \mathcal{V}=-\frac{p}{q}  \,.
	  \label{potential integrable}
	\end{equation} 
Our approach is reminiscent of inverse scattering methods: instead of starting with some potential $\mathcal V$ and determining the mass function $C$, we postulate some suitable mass function $C$, use \eqref{eq:angelinajolie} to get the functions $p$ and $q$, and obtain the potential $\mathcal V$ as output by virtue of \eqref{potential integrable}.\footnote{The analog of the integrability condition \eqref{side3} would be difficult to formulate if one started from a generic potential $\mathcal{V}$ instead of starting from equation \eqref{ede}, which justifies our approach.} 
In terms of \eqref{potential integrable}, equation \eqref{Q solution} leads to
	\begin{equation}
	Q(X) = \int^X \frac{\partial}{\partial Y} \left(\frac{p}{q}\right) \extd X' = \int^X \left( \frac{1}{q}\frac{\partial p}{\partial Y} - \frac{p}{q}~\frac{1}{q}\frac{\partial q}{\partial Y} \right) \extd X' ~.
	\end{equation}
The first term in the integral on the right can be rewritten with the relation \eqref{side3}.
	\begin{equation}
	Q(X) = \int^X \left( \frac{1}{q}\frac{\partial q}{\partial X'} + \mathcal{V}~\frac{1}{q}\frac{\partial q}{\partial Y} \right) \extd X'
	\end{equation} 
Performing the change of integration variable $\extd X=\mathcal{V}^{-1}\extd Y$ in the second term, and using $\extd q= \frac{\partial q}{\partial Y} ~ \extd Y  + \frac{\partial q}{\partial X} ~ \extd X$, we finally obtain
	\begin{equation}
	Q(X) = \int \frac{1}{q}~ \extd q = \ln\, q(X) 
	\label{Q2}
	\end{equation} 
where, as explained in the previous section, the integration constant does not contain physical information and has therefore been set to zero. %Notice the explicit dependence of $Q(X)$ in the Casimir via $\frac{\partial C}{\partial Y} = q$.
In general, $Q(X)$ depends on the Casimir $C$ when $q$ depends on $Y$, since we use the integrated version of \eqref{stage 1} to express $Y$ as function of $X$ and the Casimir $C$. Power-counting renormalizable models, $\mathcal{ V}(X,\,2 Y)=V(X)-YU(X)$, are a notable exception: for these models, the integrating factor $Q(X)$ is independent from the Casimir $C$.
	
Up to this stage, given $p$, $q$ in equation \eqref{ede} satisfying the integrability condition \eqref{side3}, we have shown how to write explicitly the line element \eqref{solution in metric} using \eqref{Q2} and the constraint on $Y$ that can be readily obtained from the Casimir conservation deduced from \eqref{ede}. Now we intend to determine the ground states of these families of metrics, which are characterized by solutions with {\it (i)}~constant curvature and {\it (ii)}~vanishing Casimir function. The first condition is by choice --- we are mostly interested in models that have a maximally symmetric ground state, like flat space, Rindler, or (A)dS$_2$. The second condition is without loss of generality: we can always shift the whole spectrum in such a way that the value of the Casimir function for the ground state vanishes.

The Ricci scalar $R$ is expressed in terms of the Killing norm $K$ as\footnote{Zeros of the Killing norm $K$ in \eqref{Killing norm} correspond to loci of Killing horizons. They arise either for $q=0$ or $Y=0$. The case of $q=0$ leads to an extremal Killing horizon, as $q=0$ is a double zero for the Killing norm. The loci of non-extremal Killing horizons are those for which $Y=0$.}
	\begin{equation}
	    R = \partial_r^2 K \qquad\qquad K =  2 q^2 Y \,.
	    \label{Killing norm}
	\end{equation} 
We distinguish three types of vacua, depending on the values of the curvature and Killing norm:
	\begin{itemize}
	    \item {\bf Minkowski ground states} are characterized by vanishing Ricci scalar, $R=0$, and constant Killing norm, $K(r)=\rm const.$
	    \item {\bf Rindler ground states} are given by vanishing Ricci scalar, $R=0$, and linear Killing norm, $K(r) \propto r$.
	    \item {\bf (A)dS ground states} are characterized by positive (negative) Ricci scalar, so that the Killing norm is quadratic. We can distinguish between Poincar\'e (A)dS with $K(r) \propto (-) r^2$, global (A)dS with $K(r) \propto ((-)r^2+1)$ and (A)dS Rindler where $K(r) \propto (-) r^2 + Ar + B$ for some constants $A$, $B$. In the following, when discussing (A)dS ground states, we will only refer to Poincar\'e (A)dS ground states. 
	\end{itemize}

\section{Three-parameter family as example}
\label{sec:4}

Power-counting renormalizable models involve potentials and Casimirs that are at most linear in $Y$. The ground states of these models have been extensively discussed in the literature (see e.g.~\cite{Katanaev:1996ni}). Here, we focus on the simplest general dilaton gravity that is not power-counting renormalizable by considering a Casimir $C(X,Y)$ quadratic in $Y$,
	\begin{equation}
	    \label{side15}
	    C(X,Y) = V_0(X) + V_1(X)\, Y + V_2(X)\, Y^2 
	\end{equation} 
where the potentials $V_0$, $V_1$, $V_2$ are taken to be monomials in $X$.
	\begin{equation}
	    V_0(X) = A~X^a \qquad\qquad V_1(X) = B~X^b \qquad\qquad V_2(X) = D~X^c  \label{3param}
	\end{equation} 
The parameters $A$, $B$, $D$ and $a$, $b$, $c$ are constant. The condition {\it (ii)} to obtain ground state solutions requires solving $C(X,Y)=0$ for $Y$. This leads to two branches of solutions.
	\begin{equation}
	    Y^{\pm} = \frac{-V_1 \pm \sqrt{V_1^2 - 4V_2V_0}}{2V_2} 
	    \label{et1}
	\end{equation}
Evaluating the function $q(X,Y)$ on these solutions yields
	\begin{equation}
	    q(X,Y^{\pm}) = V_1 + 2V_2Y_{\pm} = \pm\sqrt{V_1^2 - 4V_2V_0} ~. \label{et2}
	\end{equation}
Plugging \eqref{et1} and \eqref{et2} into the second equation of \eqref{Killing norm} gives the ground state Killing norm 
	\begin{equation}
	    	K^\pm_0:=K^{\pm}|_{C=0} = (V_1^2 - 4V_2V_0) \frac{-V_1 \pm \sqrt{V_1^2 - 4V_2V_0}}{V_2} ~.
	    	\label{vanishC}
	\end{equation} 
For the three-parameter family \eqref{3param}, this reads as
	\begin{equation}
	    \label{abcKillingnorm}
	    K^{\pm}_0 = \big(B^2 X^{2b} - 4AD~X^{a+c}\big) \Big( -\frac{B}{D} X^{b-c} \pm \sqrt{\frac{B^2}{D^2} X^{2b-2c} - 4\frac{A}{D} X^{a-c}} \Big) ~.
	\end{equation}
By expanding $Y^{\pm}$ for small values of $D$, one can see that the branch $Y^-$ is continuously connected to the linear solution, while the branch $Y^+$ is not. This constitutes a criterion to select the branch $Y^-$ rather than $Y^+$, which is what we are going to do. 

We now focus on the condition {\it (i)} defining ground states. %, which consists in finding constant curvature solutions. 
Following the classification of vacua provided at the end of the previous section, we treat each case separately. We determine first the conditions on the model parameters $a,b,c$ such that the ground state of the corresponding model has the desired properties, then write down the potential associated with the model and finally display the metric \eqref{solution in metric2} that emerges as solution of the model, for any value of the Casimir $C$. The relation between dilaton and radial coordinate is displayed in the appendix. As an abbreviation we often use the discriminant $\Delta:=B^2-4AD$.

\subsection{Minkowski ground states}
\label{sec:4.1}

Minkowski ground states are found by requiring the expression of the Killing norm \eqref{abcKillingnorm} to be constant. Rewriting \eqref{abcKillingnorm} as
    \begin{equation}
	    K^{-}_0 = X^{3b-c}\, \big(B^2 - 4AD~X^{a+c-2b}\big) \Big( -\frac{B}{D} - \sqrt{\frac{B^2}{D^2} - 4\frac{A}{D} X^{a+c-2b}} \Big) = \textrm{const} ~,
	\end{equation}
we see that this condition is satisfied for generic $A$, $B$ and $D$ only if each term in the product is constant, and therefore $a + c -2b = 0$ and $3b - c = 0$. Minkowski ground state models then have to obey the relations
	\begin{equation}
\boxed{	
\phantom{\Bigg(}
\textrm{Minkowski\;ground\;state\;models:}\qquad\qquad a = -b \qquad\qquad c = 3b
\phantom{\Bigg)}
}
\label{eq:MGS}
	\end{equation}
between the parameters, yielding the potential
	\begin{equation}
\boxed{	
\phantom{\Bigg(}
{\cal V}_{\textrm{\tiny Mink}}\big(X,\,-(\partial X)^2\big) = a\,X^{2a-1}\,\frac{4A +2 B X^{-2a}(\partial X)^2 - 3D X^{-4a} (\partial X)^4}{-4B + 4 D X^{-2a}(\partial X)^2}\,.
\phantom{\Bigg)}
}
\label{eq:MGSpot}
	\end{equation}
The metric \eqref{solution in metric2} for Minkowski ground state models simplifies to ($\Delta=B^2-4AD$)
\eq{
\extd s^2=2\extd v \extd r - \whatever\,\Big(\Delta+4CDX^{-a}\Big)\Big(B+\sqrt{\Delta+4CDX^{-a}}\Big)\,\extd v^2
}{eq:MGSmetric} 
where we labelled by $\whatever$ the freedom of fixing the integration constant in the function $Q(X)$ given in \eqref{Q2}. Demanding the terms proportional to the Casimir $C$ to be subleading for large $X$ establishes the convexity condition $a>0$. For positive $a$ and a convenient choice of $\whatever$, the metric \eqref{eq:MGSmetric} has the asymptotically flat expansion
\eq{
\extd s^2=2\extd v \extd r - \Big(1-MX^{-a} + {\cal O}\big(M^2X^{-2a}\big)\Big)\,\extd v^2
}{eq:MGSasy}
where the mass parameter $M$ is proportional to $C$. Demanding that large values of the radial coordinate also correspond to large values of the dilaton requires $a\leq 1$. Thus, standard Minkowski ground state models obey $a\in(0,1]$ in addition to \eqref{eq:MGS}.

Non-generic cases for specific values of the constants $A$, $B$, $D$ exist as well. If $A=0$ and $B\neq0$, we get the ground state condition $c=3b$ with no further restriction. In the case of $B=0$ and $A\neq0$, $K^{-}_0=\textrm{const}$ leads to $c=-3a$. The case $D=0$ leads to a two-parameter family of power-counting renormalizable models with the conditions on the parameter $K_0 = 2 q^2 Y = -2AB~X^{a+b}= \textrm{const}$, so that $a=-b$. Comparing with the well-known two-parameter family $(a'b')$ (see e.g.~\cite{Katanaev:1996ni}) where the potential reads as $\mathcal{V}_{(a'b')}(X,2Y) = \frac{a'}{X} Y +  X^{a'+b'}$ and using the dictionary $a'=-b$ and $b'=a-1$, we recover the known result for Minkowski ground state $a'=1+b'$, which constitutes a consistency check of our derivation.

\subsection{Rindler ground states}
\label{sec:4.2}

With the radial coordinate defined by $r = \int^X q(X',Y) \extd X'$, the condition $K^-_0\propto r$ reads as follows
\begin{align}
	    \nonumber
	    K^{-}_0 &= \big(B^2 X^{2b} - 4AD~X^{a+c}\big) \Big( -\frac{B}{D} X^{b-c} - \sqrt{\frac{B^2}{D^2} X^{2b-2c} - 4\frac{A}{D} X^{a-c}} \Big) \\ &\propto
	     X^{b+1}\,\Big[\frac{B}{b+1} + \frac{2D}{c+1} \Big( -\frac{B}{D} - \sqrt{\frac{B^2}{D^2} - 4\frac{A}{D} X^{a-2b+c}} \Big)\Big] ~.
\end{align} This can be rewritten as 
\begin{equation}
	\label{side10}
	\left( -\frac{B}{D} - \sqrt{\frac{B^2}{D^2} - 4\frac{A}{D} X^{a-2b+c}} \right) \left( -\frac{2D}{c+1} + X^{2b-c-1} ( B^2 - 4AD~X^{a-2b+c} ) \right) \propto  \frac{B}{b+1} ~.
\end{equation} 
For generic $A$, $B$ and $D$ this can be solved for the parameters
	\begin{equation}
\boxed{	
\phantom{\Bigg(}
\textrm{Rindler\;ground\;state\;models:}\qquad\qquad a=1\qquad\qquad
	c=2b-1
\phantom{\Bigg)}}
	\end{equation}
yielding the potential
	\begin{equation}
\boxed{	
\phantom{\Bigg(}
{\cal V}_{\textrm{\tiny R}}\big(X,-(\partial X)^2\big) =  X^{-b}\frac{4A -2 bB  X^{b-1}(\partial X)^2 + (2b-1)D  X^{2b-2}(\partial X)^4}{-4B + 4 D X^{b-1}(\partial X)^2}\,.
\phantom{\Bigg)}
}
	\end{equation}
The metric \eqref{solution in metric2} for Rindler ground state models simplifies to ($\Delta=B^2-4AD$)
\eq{
\extd s^2=2\extd v \extd r - \whatever X^{1+b}\Big(\Delta+\frac{4CD}{X}\Big)\Big(B+\sqrt{\Delta+\frac{4CD}{X}}\Big)\,\extd v^2
}{eq:RGS} 
where $\whatever$ has the same meaning as before. The terms proprotional to the Casimir $C$ are always subleading at large $X$. Demanding that large values of the radial coordinate also correspond to large values of the dilaton requires $b\geq -1$. For these standard Rindler ground state models, the metric \eqref{eq:RGS} has the asymptotically Rindler expansion
\eq{
\extd s^2=2\extd v \extd r - \Big(r-Mr^{b/(1+b)} + {\cal O}\big(M^2r^{(b-1)/(1+b)}\big)\Big)\,\extd v^2
}{eq:RGSasy}
where the mass parameter $M$ is proportional to $C$.

Non-generic cases for specific values of the constants $A$, $B$, $D$ exist as well. For $A= 0$, the condition $K^-_0 \propto r$ is satisfied provided $c=2b-1$. For $B=0$ and $A\neq0$ we get the solution $a=1$ satisfying $K^{-}_0\propto r$. For $D=0$, the condition $K_0^- \propto r$ yields $a=1$ with $b$ arbitrary. In terms of the $(a'b')$ family, we get $b'=0$, which confirms the known result \cite{Katanaev:1996ni}.

\subsection{(A)dS ground states}
\label{sec:4.3}

The (A)dS condition $K^-_0\propto r^2$ is given by
    \begin{align}
	\nonumber
	K^{-}_0 &= X^{3b-c}\,\big(B^2 - 4AD~X^{a-2b+c}\big)  \Big( -\frac{B}{D} - \sqrt{\frac{B^2}{D^2} - 4\frac{A}{D} X^{a-2b+c}} \Big) \\ 
	& \propto X^{2b+2} \left[ \left( \frac{B}{b+1} \right)^2 +\frac{4BD}{(b+1)(c+1)} \left( -\frac{B}{D} - \sqrt{\frac{B^2}{D^2} - 4 \frac{A}{D} X^{a-2b+c}} \right) + \right. \nonumber \\
	& \left. +\left( \frac{2D}{c+1} \right)^2 \left( -\frac{B}{D} - \sqrt{\frac{B^2}{D^2} - 4 \frac{A}{D} X^{a-2b+c}} \right)^2 \right] ~.
	\end{align}
For generic $A$, $B$ and $D$ this is solved for the parameters obeying
    \begin{equation}
\boxed{	
\phantom{\Bigg(}
\textrm{(A)dS\;ground\;state\;models:}\qquad\qquad
	    a = b + 2 \qquad\qquad c = b - 2\,. 
\phantom{\Bigg)}}	    
	\end{equation}
yielding the potential
	\begin{equation}
\boxed{	
{\cal V}_{\textrm{\tiny (A)dS}}\big(X,-(\partial X)^2\big) = X\frac{4aA - 2(a-2)B X^{-2}(\partial X)^2 + (a-4)D X^{-4}(\partial X)^4}{-4B + 4 D X^{-2} (\partial X)^2}\,.
}
	\end{equation}	
The metric \eqref{solution in metric2} for (A)dS ground state models simplifies to ($\Delta=B^2-4AD$)
\eq{
\extd s^2=2\extd v \extd r + \whatever X^{2a-2}\Big(\Delta+\frac{4CD}{X^a}\Big)\Big(B+\sqrt{\Delta+\frac{4CD}{X^a}}\Big)\,\extd v^2
}{eq:AGS} 
where again $\whatever$ has the same meaning as before. The terms proprotional to the Casimir $C$ are subleading at large values of the dilaton if $a>0$. If $a>1$, then large values of the radial coordinate also correspond to large values of the dilaton. For these standard (A)dS ground state models, the metric \eqref{eq:AGS} has the asymptotically (A)dS expansion
\eq{
\extd s^2=2\extd v \extd r + \Big(\Lambda\,r^2 + M\,r^{(a-2)/(a-1)} {\cal O}\big(M^2r^{-2/(a-1)}\big)\Big)\,\extd v^2
}{eq:AGSasy}
where $\Lambda=\Lambda(A,B,D)$ is positive (negative) for (A)dS, and the mass parameter $M$ is proportional to the Casimir $C$.

Again, some non-generic cases for specific values of the constants $A$, $B$, $D$ exist. If $A=0$, $K^-\propto r^2$ yields $c=b-2$. For $B=0$ we get $c=a + 4b$. For $D=0$ we get the result $a=b+2$ from the condition $K_0 \propto r^2$. This is consistent with the AdS ground states previously found for the model $\mathcal{V}_{(a')}(X,2Y) = \frac{a'}{X} Y +  X$ \cite{Katanaev:1996ni}.

\section{Generalizations}
\label{sec:6}

There are three types of generalizations of the standard Minkowski-, Rindler-, and (A)dS-ground state models discussed in this work.
\begin{enumerate}
    \item {\bf Beyond standard models.} One can relax the convexity conditions imposed as standard. For instance, one could consider Minkowski ground state models \eqref{eq:MGSpot} where the parameter $a$ lies outside the interval $(0,1]$. Then either the weak coupling region, $X\to\infty$, no longer corresponds to large values of the radial coordinate, $r\to\infty$, or the Killing norm is dominated asymptotically by the terms containing the Casimir $C$. In the latter case, generic solutions no longer are asymptotically flat. Similar remarks apply to Rindler and (A)dS ground state models. In the restricted class of power-counting renormalizable models, this issue was addressed in \cite{Grumiller:2014oha,Bagchi:2014ava}.
    \item {\bf Beyond maximally symmetric ground states.} One can give up the requirement of having a maximally symmetric ground state and study all members of the family of models defined by \eqref{side15} with \eqref{3param}. It could be interesting to scan this model space by imposing as constraint reasonable thermodynamics, along the lines of section 4.2 in \cite{Grumiller:2007ju}.
    \item {\bf Beyond quadratic ansatz.} One can extend the discussion beyond the 3-parameter family considered in the present work. In particular, generalizing the ansatz \eqref{side15} to $C(X,Y)=\sum_{n=0}^4 V_n(X) Y^n$ yields explicit (if cumbersome) expressions, and all our results generalize straightforwardly to this case. Generalizations beyond these algebraic examples require new techniques or special choices of $C(X,Y)$.
\end{enumerate}

Besides these generalizations, there are also some evident potential applications of the models studied in the present work. We list three examples for possible follow-up work and expect that some of these points will be addressed in the near future. 
\begin{enumerate}
\item {\bf Thermodynamics and holographic renormalization.} Presently, it is unclear which boundary terms have to be added to the bulk action \eqref{eq:1} to render the variational principle well-defined. For the class of power-counting renormalizable models, these boundary terms were constructed in \cite{Grumiller:2007ju} (using the Hamilton--Jacobi method, see e.g.~\cite{Papadimitriou:2010as}). They play an important role in thermodynamics since in the Euclidean version of the theory the Helmholtz free energy is determined from the (holographically renormalized) on-shell action. It could be rewarding to extend the thermodynamical discussion of \cite{Grumiller:2007ju} to generic 2d dilaton gravity models \eqref{eq:1}, like in section 5 of \cite{Grumiller:2021cwg}.
\item {\bf Asymptotic symmetries and boundary excitations.} Models with maximally symmetric ground states could allow for some standard boundary conditions, using notions of locally asymptotically flat or locally asymptotically (A)dS spacetimes, along the lines of e.g.~\cite{Grumiller:2017qao,Grumiller:2021cwg}. 
\item {\bf SYK-like correspondences.} Given the relation between Schwarzian actions and AdS isometries (see \cite{Maldacena:2016hyu} and the reviews \cite{Mertens:2018fds,Sarosi:2017ykf,Gu:2019jub}) one might expect SYK-like models \cite{Kitaev:15ur,Sachdev:1992fk} dual to the AdS ground state models. Similarly, by analogy to \cite{Afshar:2019axx} Rindler ground state models could lead to a twisted warped boundary action.
\end{enumerate}

\begin{acknowledgement}
DG thanks the organizers of the MATRIX Event \href{https://www.matrix-inst.org.au/events/2d-supersymmetric-theories-and-related-topics/}{``2D Supersymmetric Theories and Related Topics''} for their efforts.

DG and RR were supported by the Austrian Science Fund (FWF), projects P~30822, P~32581, P~33789, and Y~1447.  
\end{acknowledgement}

\section*{Appendix --- Relation between dilaton and radial coordinate}

The relation between dilaton and radial coordinate,
\eq{
\extd r = e^{Q(X)}\,\extd X
}{eq:app1}
can be integrated in closed form in terms of elementary functions and the hypergeometric function $_2F_1$ for all models considered in our work. Dropping the integration constant in $Q(X)$, we list now these results, using again the definition $\Delta=B^2-AD$ and introducing the definition $M=-4CD/\Delta$.

For Minkowski ground state models, we obtain
\begin{multline}
r = \frac{2X\sqrt{\Delta(1-M X^{-a})}}{(a-2)(3a-2)M^2}\,\Big(\frac{2a}{a+2}\,X^a\,_2F_1(1,1+\tfrac1a,\tfrac32+\tfrac1a;\,X^a/M)\\
+M\big(a+(2-a)MX^{-a}\big)\Big)
\label{eq:app2}
\end{multline}
which for large values of $X$ and $a\in(0,1]$ expands as
\eq{
r \propto X^{1-a} + \dots
}{eq:app3}

For Rindler ground state models, we obtain
\eq{
r = \frac23\,\sqrt{\Delta}\,\Big(\frac{1}{M}\Big)^{1/2-b}\,(X-M)^{3/2}\,_2F_1(\tfrac32,\tfrac12-b,\tfrac52;\,1-X/M) 
}{eq:app4}
which for large values of $X$ and $b>-1$ expands as
\eq{
r \propto X^{1+b} + \dots
}{eq:app5}

For (A)dS ground state models, we obtain
\eq{
r = \frac{a\sqrt{\Delta(1-MX^{-a})}X^{a-1}}{(a-2)(a-1)}\,\Big(\,_2F_1(1,1-\tfrac1a,\tfrac32-\tfrac1a;\,X^a/M)+1-\tfrac2a\Big)
}{eq:app6}
which for large values of $X$ and $a>1$ expands as
\eq{
r \propto X^{-1+a} + \dots
}{eq:app7}

In all cases, the omitted proportionality constants are independent from $M$ and can be adjusted conveniently by fixing the integration constant in $Q(X)$.

%\bibliographystyle{spmpsci}
%\bibliography{review} 

\end{document}